\documentclass[prd,superscriptaddress,showpacs,floats,nofootinbib,preprint,floatfix]{revtex4}
\usepackage{layout}
\usepackage{amsmath}
\usepackage{textcomp}
\usepackage{hyperref}

\usepackage{latexsym}
\usepackage{bbm}
\usepackage{bbold}
\usepackage{slashed}
\newcommand{\beq}{\begin{equation}}
\newcommand{\eeq}{\end{equation}}
\newcommand{\beqa}{\begin{eqnarray}}
\newcommand{\eeqa}{\end{eqnarray}}
\newcommand{\bold}{\boldsymbol}

\begin{document}

\title{Meson Correlation Function and Screening Mass in Thermal QCD}
\author{Piotr Czerski\email{Piotr.Czerski@ifj.edu.pl}}
\affiliation{The H. Niewodnicza\'nski Institute of Nuclear Physics, Polish Academy of Sciences,\\
     ul. Radzikowskiego 152, 31-342 Krak\'ow, Poland}
\begin{abstract}
{Analytical results for the spatial dependence of the correlation functions for all meson excitations in perturbative Quantum Chromodynamics, the lowest order, are calculated. The meson screening mass is
obtained as a large distance limit of the correlation function.
Our analysis leads to a better understanding of the excitations of Quark Gluon Plasma 
at sufficiently large temperatures and may be of relevance 
for future numerical calculations with fully interacting Quantum Chromodynamics.}
\end{abstract}

\maketitle

\section{Introduction}
In this work we study mesonic correlation functions in high temperature Quantum Chromodynamics (QCD) in the lowest
order of the perturbative expansion.
We follow the method developed by \cite{ff} where the simplest pseudoscalar
case was calculated. Calculations are performed analytically with a special attention to the divergent parts, which need to be regularized and we adopt a Pauli-Villars \cite{zuber}
regularization scheme.
The range of the correlation function is determined by the screening mass, which can be defined as the inverse screening length characterizing
the exponential fall-off of the mesonic spatial correlator.
For zero quark mass at the high temperature $T$ the meson and baryon screening masses
 approach their ideal gas value, $2 \pi T$ and $3 \pi T$ \cite{ele,shur} respectively.

The pseudoscalar meson screening mass was calculated \cite{ff} in the noninteracting case
$m_{scr}=2 \sqrt{\pi^2 T^2+m^2}$, for finite quark mass $m$. Calculations in  the interacting Quark Gluon Plasma (QGP) 
 were done
in nonrelativistic QCD \cite{lai,vep,lai2} for all meson channels
 and in the Hard Thermal Loop Approximation (HTL) of the QCD \cite{alb} for
 the pseudo scalar one.
The last, numerical approach, requires the analytical form of the mesonic spatial 
correlation functions in the free case in order to calculate the divergent integrals.

The main scope of this work is to provide the analytical expressions for the
correlation functions for all meson channels to allow for future calculations
of the screening masses in the interacting QGP.

The importance of such evaluation  in order 
to identify the relevant degrees of freedom in hot and dense QCD was first 
pointed out in~\cite{DeTar1,DeTar2,got}, where the screening masses of mesons 
and nucleons were calculated on the lattice. More recent  results 
can be found in~\cite{mtc,taro1,taro2,wis,gavai1,gavai2,gavai3,muk1,muk2,falc1,falc2,cheng}.
 Though the evaluation of hadronic screening masses is a major achievement of  
lattice studies of the degrees of freedom characterizing the hot QCD 
(the large number of lattice sites available along the spatial directions allows 
one to study the large distance behaviour of the correlators), analytical approaches are 
not so common in the literature.
\section{Mesonic spatial correlation functions}
The lowest order quark loop contribution to the correlation function is defined by the expression
\beq
\label{gq}
G_M(q) = - 2 N_c  \hbox{T} \sum_n \int {d^3 p \over (2\pi)^3} \hbox{Tr} 
\left[ \Gamma_M \frac{i}{\slashed p + \slashed q -m} \Gamma_M \frac{i}{\slashed p -m}\right]
\eeq
from which we calculate spatial dependence for static plane-like perturbation
\beq
\label{ft}
G_M(z) =
{1 \over 2 \pi} \int\limits_{-\infty}^{+\infty} dq_z \, G_M(0,q_z^2)\, e^{iq_z z}.
\eeq
Here M=S, PS, V, PV corresponds to the scalar, pseudoscalar,
vector and pseudovector
channel ($\Gamma_S=\mathbb{1}$, $\Gamma_{PS}=\gamma_5$, $\Gamma_{V}=\gamma_{\mu}$, $\Gamma_{PV}=\gamma_5 \gamma_{\mu}$), $q^{\mu}=(i\omega_m, {\bf q}\,)$ is the external
momentum, $\omega_m=2m\pi i T$ is a boson Matsubara frequency. Summation runs over fermion Matsubara frequencies of quark internal momentum $p^{\mu}=(i\omega_n, {\bf p}\,)$, $\omega_n=(2n+1)\pi i T$ and \hbox{Tr} denotes
the trace over the spinor indices. 
\beqa
\label{tr}
\hbox{Tr} \left[ \Gamma_{M} \frac{i}{\slashed p + \slashed q -m} \Gamma_M \frac{i}{\slashed p -m} \right] &=& \\
(Scalar) & = & -\frac{4 \left(m^2+p 
   q+p^2\right)}{\left(p^2-m
   ^2\right)
   \left((p+q)^2-m^2\right)} , \nonumber \\
(Pseudo scalar) & = &
-\frac{4 \left(m^2-p 
   q-p^2\right)}{\left(p^2-m
   ^2\right)
   \left((p+q)^2-m^2\right)} , \nonumber \\
(Vector) & = &
-\frac{4 \left(4 m^2-2 p 
   q-2
   p^2\right)}{\left(p^2-m^2
   \right) \left((p+q)^2-m^2\right)} , \nonumber \\
(Pseudo vector) & = &
-\frac{4 \left(-4 m^2-2 p 
   q-2
   p^2\right)}{\left(p^2-m^2
   \right) \left((p+q)^2-m^2\right)} . \nonumber
\eeqa
We convert the sum over the frequencies \cite{kapusta} to two integrals. The first one,
 independent
of $T$ which is called a vacuum part (divergent one) and the second, $T$-dependent,
called  the matter part.
\beq
G_M = G_M^{vac}+ G_M^{mat} .
\eeq
Let's start from the vacuum part first
\beqa
\label{vac}
G_{S}^{vac}(q) &=& -8 i N_c  \int\limits_{-i\infty}^{+i\infty} \frac{dp^0}{2 \pi} 
\int \frac{d^3p}{(2 \pi)^3} \frac{p^2-q^2/4+m^2}{[(p-q/2)^2-m^2][(p+q/2)^2-m^2]} ,
\nonumber \\
G_{PS}^{vac}(q) &=& 8 i N_c  \int\limits_{-i\infty}^{+i\infty} \frac{dp^0}{2 \pi} 
\int \frac{d^3p}{(2 \pi)^3} \frac{p^2-q^2/4-m^2}{[(p-q/2)^2-m^2][(p+q/2)^2-m^2]} ,
\nonumber \\
G_{V}^{vac}(q) &=& 16 i N_c  \int\limits_{-i\infty}^{+i\infty} \frac{dp^0}{2 \pi} 
\int \frac{d^3p}{(2 \pi)^3} \frac{p^2-q^2/4-2 m^2}{[(p-q/2)^2-m^2][(p+q/2)^2-m^2]} ,
\nonumber \\
G_{PV}^{vac}(q) &=& 16 i N_c  \int\limits_{-i\infty}^{+i\infty} \frac{dp^0}{2 \pi} 
\int \frac{d^3p}{(2 \pi)^3} \frac{p^2-q^2/4+2 m^2}{[(p-q/2)^2-m^2][(p+q/2)^2-m^2]} .
\eeqa
After a Wick rotation ($p^0 \to i p_4$ and $q^0 \to i q_4$) in  the 
Euclidean metric ($q_E^2=q_4^2+\bold q^2=-q^2$) we can express (\ref{vac}) with two integrals
\beq
\label{i1}
I_{1}(m) = 8N_c \int {d^4 p_E \over (2\pi)^4} {1 \over p_E^2 + m^2},
\eeq
\beq
\label{i2}
I_{2}(m,q_E) = -4N_c \int {d^4 p_E \over (2\pi)^4} {1 \over
[(p_E-q_E/2)^2+m^2][(p_E+q_E/2)^2+m^2]},
\eeq
where $d^4 p_E = d^3 p dp_4$. Both integrals are divergent and must be regularized.
Following \cite{ff} we use Pauli-Villars regularization and define regularized functions:
\beq
\label{IR1}
I_1^R = \sum_{i=0}^{N} a_i I_1(M_i) \ \ \ \mbox{and} \ \ \ I_2^R(q_E) = \sum_{i=0}^{N} a_i I_2(M_i,q_E) ,
\eeq
where $a_0=1$, $M_0=m$ and higher masses are large. Coefficients $a_i$ for $i>0$ fulfill the
following equations
\beq
\sum_{i=0}^{N} a_i=0,\ \ \sum_{i=0}^{N}a_i M_i^2=0,\ \ ... \ \ 
\sum_{i=0}^{N}a_i M_i^{2(N-1)}=0 .
\eeq
After some calculations we have
\beqa
\label{IRR1}
I_1^R &=& \frac{N_c}{2 \pi^2}\sum_{i=0}^{N} a_i M_i^2 { \rm ln}M_i^2  ,
\\
\nonumber
I_2^R(q_E) &=& \frac{N_c}{2 \pi^2}\sum_{i=0}^{N} a_i \left[ \frac{2 M_i}{q_E}
\sqrt{1+\left(\frac{q_E}{2 M_i}\right)^2} { \rm ln} 
\left( \sqrt{1+\left(\frac{q_E}{2 M_i}\right)^2}  +\frac{q_E}{2 M_i} \right) 
+ { \rm ln}M_i \right] .
\eeqa
The vacuum part of the correlation function can be written
\beqa
\label{sq}
G_{S}^{vac}(q) &=& -I_1^R - q_E^2 I_2^R(q_E)-4 m^2 I_2^R(q_E) ,\nonumber \\
G_{PS}^{vac}(q) &=& I_1^R + q_E^2 I_2^R(q_E) , \\
G_{V}^{vac}(q) &=& 2 I_1^R +2 q_E^2 I_2^R(q_E)- 4 m^2 I_2^R(q_E) ,\nonumber \\
G_{PV}^{vac}(q) &=& 2 I_1^R + 2 q_E^2 I_2^R(q_E)+12 m^2 I_2^R(q_E) .\nonumber 
\eeqa
The next step is to obtain a $z$-dependent correlation function. Performing the Fourier
transforms of  (\ref{sq}) we find that in the complex $q_E$ plane the function $I_2^R(q_E)$
has cuts for imaginary $q_E=ik$  from $\pm 2 M_i$ to $\pm \infty$ respectively.
Performing the integration (there is no contribution from $I_1^R$ integral) the
$q_E^2 I_2^R$ gives
\beq
\label{GPS}
\frac{N_c}{4 \pi^2} \sum_{i=0}^{N} a_i  \int\limits_{2 M_i}^{\infty} dk k
\sqrt{k^2-4 M_i^2}  \ e^{-k z} = \frac{N_c}{\pi^2 z} \sum_{i=0}^{N} a_i M_i^2 K_2(2 M_i z),
\eeq
where $K_2$ is a modified Bessel function of a second kind. The $I_2^R$ gives
\beqa
\label{i2z}
I_2^R(z)&=& -\frac{N_c}{4 \pi^2} \sum_{i=0}^{N} a_i  \int\limits_{2 M_i}^{\infty} dk 
\frac{\sqrt{k^2-4 M_i^2}}{k}  \ e^{-k z} \\
&=& - \frac{N_c }{8 \pi^2} \sum_{i=0}^{N} a_i M_i G_{1,3}^{3,0}\left(z^2 M^2_i  \left|
\begin{array}{c}
 1 \\
 -\frac{1}{2},0,\frac{1}{2}
\end{array} \right.
\right),\nonumber
\eeqa
where $G_{1,3}^{3,0}$  is a Meijer G function \cite{meijer}. In both summations 
(\ref{GPS},\ref{i2z}) the only first term survives and finally
\beqa
\label{sq1}
G_{S}^{vac}(z) &=& -\frac{N_c}{\pi^2 z} m^2 K_2(2 m z)+ \frac{N_c}{2 \pi^2} m^3
\mathcal{M}^{vac}(m,z) , \nonumber \\
G_{PS}^{vac}(z) &=&   \frac{N_c}{\pi^2 z} m^2 K_2(2 m z) ,\\
G_{V}^{vac}(z) &=&  \frac{2 N_c}{\pi^2 z} m^2 K_2(2 m z)  + \frac{N_c}{2 \pi^2} m^3
\mathcal{M}^{vac}(m,z) ,\nonumber \\
G_{PV}^{vac}(z) &=&  \frac{ 2 N_c}{\pi^2 z} m^2 K_2(2 m z)  -  \frac{3 N_c}{2 \pi^2} m^3
\mathcal{M}^{vac}(m,z) , \nonumber 
\eeqa
where $\mathcal{M}^{vac}(m,z)=G_{1,3}^{3,0}\left(z^2 m^2  \left|
\begin{array}{c}
 1 \\
 -\frac{1}{2},0,\frac{1}{2}
\end{array} \right.
\right)$.
The matter part of the correlation function is
\beq
\label{matt1}
G_{M}^{matt}(q) = -4 i N_c  \int\limits_{-i\infty+\epsilon}^{+i\infty+\epsilon} \frac{dp^0}{2 \pi} 
\int \frac{d^3p}{(2 \pi)^3} \frac{\hbox{Tr} \left[ \Gamma_{M} \frac{i}{\slashed p + \slashed q -m} \Gamma_M \frac{i}{\slashed p -m} \right]}{e^{{p_0}/{T}}+1} ,
\eeq
where Tr is evaluated in (\ref{tr}). After change of variables $p \to p'-q/2$, like in the
vacuum part, and keeping the static limit ($q^0=0$) we evaluate the energy integral by deforming a contour of integration around the poles
\beq
p^0=\omega_{\pm}(\bold p,\bold q)=\sqrt{m^2+(\bold p \pm \bold q /2)^2}
\eeq
and obtain
\beqa
\label{matt}
G_{S}^{matt}(q) &=& -16 N_c  
\int \frac{d^3p}{(2 \pi)^3} \left[  \frac{\bold p \bold q +q^2/2+2 m^2}{4 \omega_+ \bold p \bold q
(e^{\omega_+/T}+1)}    +  \frac{\bold p \bold q -q^2/2-2 m^2}{4 \omega_- \bold p \bold q
(e^{\omega_-/T}+1)}      \right] ,
\nonumber \\
G_{PS}^{matt}(q) &=& +16 N_c  
\int \frac{d^3p}{(2 \pi)^3} \left[  \frac{\bold p \bold q +q^2/2}{4 \omega_+ \bold p \bold q
(e^{\omega_+/T}+1)}    +  \frac{\bold p \bold q -q^2/2}{4 \omega_- \bold p \bold q
(e^{\omega_-/T}+1)}      \right] 
 \\
G_{V}^{matt}(q) &=& -32 N_c  
\int \frac{d^3p}{(2 \pi)^3} \left[  \frac{\bold p \bold q +q^2/2-m^2}{4 \omega_+ \bold p \bold q
(e^{\omega_+/T}+1)}    +  \frac{\bold p \bold q -q^2/2+m^2}{4 \omega_- \bold p \bold q
(e^{\omega_-/T}+1)}      \right] ,
\nonumber \\
G_{PV}^{matt}(q) &=& -32 N_c  
\int \frac{d^3p}{(2 \pi)^3} \left[  \frac{\bold p \bold q +q^2/2+3 m^2}{4 \omega_+ \bold p \bold q
(e^{\omega_+/T}+1)}    +  \frac{\bold p \bold q -q^2/2-3 m^2}{4 \omega_- \bold p \bold q
(e^{\omega_-/T}+1)}      \right] . \nonumber
\eeqa
Then we simplify by changing the 
variables $\bold p' \to \bold p + \bold q/2$ in the first 
part of the sum and $\bold p' \to \bold p - \bold q/2$ in the second part and integrate over angles.
\beqa
\label{matta}
G_{S}^{matt}(q) &=&  \frac{N_c}{\pi^2}  \int\limits_{0}^{\infty}
dp p^2  \frac{1}{\omega_p}  \frac{1}{e^{\omega_p/T}+1} \left\{ A_1(p,q)+2 m^2 A_2(p,q) \right\}
 ,\nonumber \\
G_{PS}^{matt}(q) &=&  - \frac{N_c}{\pi^2}  \int\limits_{0}^{\infty}
dp p^2  \frac{1}{\omega_p}  \frac{1}{e^{\omega_p/T}+1} \left\{A_1(p,q)\right\}
, \\
G_{V}^{matt}(q) &=& - \frac{2 N_c}{\pi^2}  \int\limits_{0}^{\infty}
dp p^2  \frac{1}{\omega_p}  \frac{1}{e^{\omega_p/T}+1} \left\{ A_1(p,q)- m^2 A_2(p,q) \right\}
,\nonumber \\
G_{PS}^{matt}(q) &=&  - \frac{2 N_c}{\pi^2}  \int\limits_{0}^{\infty}
dp p^2  \frac{1}{\omega_p}  \frac{1}{e^{\omega_p/T}+1} \left\{ A_1(p,q)+3 m^2 A_2(p,q) \right\}
, \nonumber 
\eeqa
where $\omega_p=\sqrt{p^2+m^2}$, $p= | \bold p|$, $q= | \bold q|$  and
\beqa
\label{A}
A_1(p,q) &=& 4 +\frac{q}{p} \ln  \left| \frac{2 p -q}{2 p +q} \right| ,\nonumber \\
A_2(p,q) &=& \frac{2}{p q} \ln  \left| \frac{2 p -q}{2 p +q} \right| .
\eeqa
Since the whole $q$ dependence is in the $A_1$ and $A_2$ functions, 
we only need the Fourier transform, see
 (\ref{ft}), of $A_1$ and $A_2$
\beqa
\label{Az}
A_1(p,z) &=& {1 \over 2 \pi} \int\limits_{-\infty}^{+\infty} dq \,
 \left\{ 4 +\frac{q}{p} \ln  \left| \frac{2 p -q}{2 p +q} \right| \right\} \, e^{iq z} , \nonumber \\
A_2(p,z) &=& {1 \over 2 \pi} \int\limits_{-\infty}^{+\infty} dq \,
\frac{2}{p q} \ln  \left| \frac{2 p -q}{2 p +q} \right| \, e^{iq z} .
\eeqa
In order to perform the integration we use the identity
\beq
\label{log}
\ln  \left| \frac{2 p -q}{2 p +q} \right|=
\lim_{\epsilon \to 0} \frac{1}{2} \ln   \frac{(2 p -q)^2+\epsilon^2}{(2 p +q)^2+\epsilon^2} .
\eeq
The function (\ref{log})  has two cuts on the upper half plane $q=ik+2p$ and $q=ik-2p$.  Fourier transform of  $A_1$ is easy to obtain
\beqa
\label{A1z}
A_1(p,z) &=& \lim_{\epsilon \to 0} {1 \over 2 \pi} \int\limits_{-\infty}^{+\infty} dq \,
 \left\{ 4 +\frac{q}{2p} \ln  \frac{(2 p -q)^2+\epsilon^2}{(2 p +q)^2+\epsilon^2}
 \right\} \, e^{iq z}  \nonumber \\
 &=& \frac{\sin (2pz)}{p z^2} - \frac{2 \cos (2 p z)}{z} .
\eeqa
The second case, $A_2$, is more complicated and we finish with two integrals
\beqa
\label{A2z}
A_2(p,z) &=& \lim_{\epsilon \to 0} {1 \over 2 \pi} \int\limits_{-\infty}^{+\infty} dq \,
 \frac{1}{pq} \ln  \frac{(2 p -q)^2+\epsilon^2}{(2 p +q)^2+\epsilon^2}
  \, e^{iq z}  \\
 &=& - \lim_{\epsilon \to 0} {1 \over p} \left[ 
  e^{2 i p z} \int\limits_{\epsilon}^{\infty} \frac{dk}{ik+2p} \, \,  e^{-kz}
- e^{-2 i p z} \int\limits_{\epsilon}^{\infty} \frac{dk}{ik-2p} \, \,  e^{-kz}
 \right] , \nonumber
\eeqa
where
\beq
\label{ipm}
\lim_{\epsilon \to 0} \int\limits_{\epsilon}^{\infty} \frac{dk}{ik\pm 2p} \,  e^{-kz} 
= - i e^{\mp 2 i p z} \left[ \Gamma(0,\mp 2ipz)+\ln \left(\frac{\pm i}{2p}\right) 
-\ln(z)+\ln(\mp i p z) \right]
\eeq
so that finally we have
\beq
\label{A2zz}
A_2(p,z) = \frac{i}{p}  \left[ \Gamma(0,- 2ipz)- \Gamma(0,2 ipz) \right] ,
\eeq
where $\Gamma(0,ix)$ is the incomplete gamma function.  In order to obtain the
 $z$-dependent
correlation function in the matter case we have to insert (\ref{A1z},\ref{A2z}) into
(\ref{matta}). To perform integration over $p$ we can use the identity
\beq
\label{exp}
 \frac{1}{e^{\omega_p/T}+1} = \frac{1}{2} - \sum_{l=-\infty}^{\infty} 
\frac{\omega_p T}{(2l+1)^2 \pi^2 T^2+\omega_p^2} .
\eeq
We then obtain
\beqa
\label{a1}
\int\limits_{0}^{\infty}   \frac{p^2 dp}{\omega_p} \left[ \frac{1}{2} - \sum_{l=-\infty}^{\infty} \nonumber
\frac{\omega_p T}{(2l+1)^2 \pi^2 T^2+\omega_p^2} \right] A_1(p,z) = 
\frac{m^2}{z} K_2(2mz)- \\
- \frac{\pi T}{2 z^2} \sum_{l=-\infty}^{\infty}  \nonumber
e^{-2z\sqrt{(2l+1)^2 \pi^2 T^2+m^2}}\left( 2z\sqrt{(2l+1)^2 \pi^2 T^2+m^2}+1\right)=\\
= \frac{m^2}{z} K_2(2mz)- A_1^T(p,z) ,
\eeqa
for $A_2$ the first term of the expansion  (\ref{exp}) gives
\beq
\label{a2}
\int\limits_{0}^{\infty}   \frac{p^2 dp}{\omega_p}  \frac{1}{2}  A_2(p,z) = 
-\frac{1}{4 \pi z} \Re \left[
G_{2,4}^{4,1}\left(- z^2 M^2  \left|
\begin{array}{c}
 \frac{1}{2},\frac{3}{2} \\
 0,\frac{1}{2},\frac{1}{2},1
\end{array} \right.
\right) \right] = -\frac{1}{4 \pi z} \mathcal{M}^{matt}(m,z) .
\eeq
For each element of the sum  (\ref{exp}) (for every $l$ and $T$) we have
\beq
\label{at}
\int\limits_{0}^{\infty}   \frac{ T  p^2 dp}{(2l+1)^2 \pi^2 T^2+\omega_p^2}   A_2(p,z) = 0 .
\eeq
Collecting (\ref{a1},\ref{a2},\ref{at}) and  (\ref{matta}) we have
\beqa
\label{mattz}
G_{S}^{matt}(z) &=&  \frac{N_c m^2}{\pi^2 z}  K_2(2 m z) - \frac{N_c}{\pi^2} A_1^T(p,z)
- \frac{N_c m^2}{2 \pi^3 z} \mathcal{M}^{matt}(m,z) ,
\nonumber \\
G_{PS}^{matt}(z) &=&  - \frac{N_c m^2}{\pi^2 z}  K_2(2 m z) + \frac{N_c}{\pi^2} A_1^T(p,z) ,
 \\
G_{V}^{matt}(z) &=& - \frac{2N_c m^2}{\pi^2 z}  K_2(2 m z) + \frac{2N_c}{\pi^2} A_1^T(p,z)
- \frac{N_c m^2}{2 \pi^3 z} \mathcal{M}^{matt}(m,z) ,
\nonumber \\
G_{PS}^{matt}(z) &=& - \frac{2N_c m^2}{\pi^2 z}  K_2(2 m z) + \frac{2N_c}{\pi^2} A_1^T(p,z)
+ \frac{3 N_c m^2}{2 \pi^3 z} \mathcal{M}^{matt}(m,z) .
\nonumber 
\eeqa
Taking into account the relation
\beq
\Re \left[ G_{2,4}^{4,1}\left(- x^2  \left|
\begin{array}{c}
 \frac{1}{2},\frac{3}{2} \\
 0,\frac{1}{2},\frac{1}{2},1
\end{array} \right.
\right) \right] = \pi x 
G_{1,3}^{3,0}\left(x^2   \left|
\begin{array}{c}
 1 \\
 -\frac{1}{2},0,\frac{1}{2}
\end{array} \right.
\right)
\eeq
we have
\beq
\mathcal{M}^{matt}(m,z) = \pi m z \mathcal{M}^{vac}(m,z)
\eeq
and finally after some cancellations
\beqa
\label{all}
G_{S}(z) &=&   \frac{-N_c T}{2 \pi z^2} \sum_{l=-\infty}^{\infty}  \nonumber
e^{-2z\sqrt{(2l+1)^2 \pi^2 T^2+m^2}}\left( 2z\sqrt{(2l+1)^2 \pi^2 T^2+m^2}+1\right)
 \\
G_{PS}(z) &=&  \frac{N_c T}{2 \pi z^2} \sum_{l=-\infty}^{\infty}  
e^{-2z\sqrt{(2l+1)^2 \pi^2 T^2+m^2}}\left( 2z\sqrt{(2l+1)^2 \pi^2 T^2+m^2}+1\right)
 \\
G_{V}(z) &=&  \frac{N_c T}{ \pi z^2} \sum_{l=-\infty}^{\infty}  
e^{-2z\sqrt{(2l+1)^2 \pi^2 T^2+m^2}}\left( 2z\sqrt{(2l+1)^2 \pi^2 T^2+m^2}+1\right)
\nonumber \\
G_{VS}(z) &=& \frac{N_c T}{ \pi z^2} \sum_{l=-\infty}^{\infty} 
e^{-2z\sqrt{(2l+1)^2 \pi^2 T^2+m^2}}\left(2z\sqrt{(2l+1)^2 \pi^2 T^2+m^2}+1\right)
\nonumber 
\eeqa
The main contribution to the summations comes from two elements $l=0$ and $l=-1$.
In the $z \to \infty$ limit all channels give the same asymptotic
\beq
G(z) \xrightarrow{z \to \infty} \frac{const}{z} e^{-2 \sqrt{\pi^2 T^2+m^2} \ z} = 
\frac{const}{z} e^{-m_{scr} \ z}
\eeq
where we identified the screening mass  $m_{scr}$
\beq
m_{scr} = 2 \sqrt{\pi^2 T^2+m^2} ,
\eeq
which for massless quarks has a simple form $m_{scr} = 2 \pi T $.
\section{Conclusions}
In this work we have shown that due to the specific cancellations of Meijer G functions
the final formulas for the spatial correlation function are compact and differ only with the sign or some numerical constants between different channels in the weakly interacting QGP.

The screening mass is the quantity which governs the large-distance exponential decay of the
correlations of mesonic current operators and is of course equal, in our system, for all mesons
and has a value $m_{scr} = 2 \sqrt{\pi^2 T^2+m^2} $, where $m$ is the free quark mass. 

We have explicitly evaluated the analytical form of the spatial,
$z$-dependent mesonic correlation functions for all mesonic channels, which is of great importance for future numerical calculations of the effective masses of mesons in the fully interacting QGP, especially that in the literature there is a discrepancy between
lattice \cite{mtc,taro1,taro2,wis,gavai1,gavai2,gavai3,muk1,muk2,falc1,falc2,cheng}
 and non-lattice \cite{lai,vep,alb} calculations of the mesonic screening masses.
In the lattice calculations in the high temperature limit, screening masses approach the non
interacting value from below, whereas in non lattice calculations they approach from above.

In papers \cite{lai,vep} the 
 analytic predictions for the screening masses,
related to various quark-antiquark excitations at high temperatures, are determined in the
nonrelativistic 3-dimensional QCD effective theory with the next-to-leading-order perturbative corrections. In paper \cite{alb} the numerical calculation in the HTL approximation,
in the pseudo scalar channel only, was performed with the help of a new technique of numerical regularization of divergent integrals with the usage of the analytical formulas of the spatial correlation functions from the non interacting case, which we provide in this work.

\end{document}